\documentclass[iop,twocolumn,numberedappendix,twocolappendix,appendixfloats]{emulateapj}
\usepackage{graphicx}
\usepackage{subfig}
\usepackage{caption}
\usepackage{multirow}

\newcommand{\Hersc}{{\it Herschel}}

\shorttitle{A Cool Dust Factory in the Crab Nebula}
\shortauthors{Gomez et al.}

\begin{document}
\title{A Cool Dust Factory in the Crab Nebula: a {\it Herschel\footnotemark[*]} study of the filaments} 

\author{H.\,L.\ Gomez\altaffilmark{1},  O.  Krause\altaffilmark{2},
  M.\,J.\,Barlow\altaffilmark{3}, B.\,M.\,Swinyard\altaffilmark{3,4},
  P. J. Owen\altaffilmark{3}, C.\,J.\,R.\,Clark\altaffilmark{1},
  M. Matsuura\altaffilmark{3}, E. L. Gomez\altaffilmark{1,5}, J. Rho\altaffilmark{6},
 M.-A. Besel\altaffilmark{2},  J. Bouwman\altaffilmark{2},
 W.\,K.\,Gear\altaffilmark{1}, Th. Henning\altaffilmark{2}, R.\,J.\
 Ivison\altaffilmark{7,8}, E. T. Polehampton\altaffilmark{4,9}, and B.\ Sibthorpe\altaffilmark{7}} 
\altaffiltext{1}{School of Physics \& Astronomy, Cardiff University,
  The Parade, Cardiff, CF24 3AA, UK}
\altaffiltext{2}{Max-Planck-Institut f\"{u}r Astronomie, K\"{o}nigstuhl 17, D-69117  Heidelberg, Germany}
\altaffiltext{3}{Deptartment of Physics and Astronomy, University College
  London, Gower Street, London WC1E 6BT, UK}
\altaffiltext{4}{Space Science and Technology Department, Rutherford
  Appleton Laboratory, Oxfordshire, OX11 0QX, UK}
\altaffiltext{5}{Las Cumbres Observatory Global Telescope Network,
  6740 Cortona Drive Suite 102, Goleta, CA 93117, US}
\altaffiltext{6}{SOFIA Science Center, Universities Space Research Association, NASA Ames Research Center, MS 232, Moffett Field, CA 94035, USA}
\altaffiltext{7}{UK Astronomy Technology Centre, Royal Observatory Edinburgh, Blackford Hill, Edinburgh EH9 3HJ, UK}
\altaffiltext{8}{Institute for Astronomy, University of Edinburgh, Blackford Hill, Edinburgh, EH9 3HJ, UK}
\altaffiltext{9}{Institute for Space Imaging Science, University of Lethbridge,
Lethbridge, Alberta, T1J 1B1, Canada}

\begin{abstract}
  Whether supernovae are major sources of dust in galaxies
  is a long-standing debate.  We present infrared
  and submillimeter photometry and spectroscopy from the {\it Herschel
    Space Observatory} of the Crab Nebula between 51 and 670\,$\mu$m
  as part of the Mass Loss from Evolved StarS program.  We
  compare the emission detected with \Hersc~with multiwavelength
  data including millimeter, radio, mid-infrared and
  archive optical images.  We carefully remove the synchrotron component using
  the \Hersc~and {\it Planck} fluxes measured in the same epoch.  The
  contribution from line emission is removed using \Hersc~spectroscopy
  combined with {\it Infrared Space Observatory} archive
  data. Several forbidden lines of carbon, oxygen and nitrogen are
  detected where multiple velocity components are resolved, deduced to
  be from the nitrogen-depleted, carbon-rich ejecta.  No spectral
  lines are detected in the SPIRE wavebands; in the PACS bands, the
  line contribution is 5\% and 10\% at 70 and 100\,$\mu$m and negligible
  at 160\,$\mu$m.  After subtracting the synchrotron and line
  emission, the remaining far-infrared continuum can be fit with
  two dust components.  Assuming standard interstellar silicates, the
  mass of the cooler component is $0.24^{+0.32}_{-0.08}\,
  M_{\odot}$ for $\rm T = 28.1^{+5.5}_{-3.2}\,K$.  Amorphous
  carbon grains require $0.11\pm 0.01\,M_{\odot}$ of dust with
  $\rm T = 33.8^{+2.3}_{-1.8}\,K$. A single temperature
    modified blackbody with $0.14\,M_{\odot}$ and $0.08\,M_{\odot}$ for silicate
    and carbon dust respectively, provides an adequate fit to the
    far-infrared region of the spectral energy distribution but is a poor fit at 24-500\,$\mu$m. The
    Crab
  Nebula has condensed most of the relevant refractory elements
  into dust, suggesting the formation of dust in core-collapse
  supernova ejecta is efficient.
\end{abstract}


\footnotetext[*]{\textit{H\lowercase{erschel}} \lowercase{is an} ESA \lowercase{space observatory with science instruments provided by} E\lowercase{uropean-led} P\lowercase{rincipal} I\lowercase{nvestigator consortia and with important participation from} NASA.}

\section{Introduction}
\label{sec:intro}
  
In galaxies, the major dust source has in the past been presumed to be
low-intermediate mass stars during their asymptotic giant branch (AGB)
phase, but when accounting for dust destruction timescales (e.g. Jones
2001; Draine 2009) and the observed total dust masses, the required
dust injection rate from stars can be an order of magnitude higher
than observed (e.g. Matsuura et al.\ 2009; Gall et al. 2011; Dunne et
al. 2011; Rowlands et al. 2012). An alternative source of dust is
required to make up the dust budget (Pipino et al. 2011; Dunne et
al. 2011). This shortfall in the dust mass estimated from AGB stars is
also observed in dusty high-redshift galaxies where the timescales for
dust production are close to, or shorter than, the lifetime of a
typical low-mass AGB star (Morgan \& Edmunds 2003; Dwek et al. 2007).

Significant dust production in supernova (SN) ejecta would alleviate
these dust budgetary problems.  SNe have long been proposed as a
source of dust (e.g. Dwek \& Scalo 1980; Clayton et al.\ 2001) yet
subsequent mid and far-infrared (FIR) observations have detected only
small quantities of warm dust in young ejecta (Sugerman et al.\ 2006;
Meikle et al.\ 2011; Kotak et al.\ 2009; Andrews et al.\ 2011; Fabbri
et al.\ 2011) and old remnants (Williams et al. 2006; Rho et
al.\ 2008).  These observed dust masses are orders of magnitude lower
than required.

In the era of the {\it Herschel Space Observatory} (Pilbratt et al.\
2010), we are now piecing together the relative contribution of
stellar sources to the dust budget in galaxies, yet dust yields from
the limited FIR studies of core-collapse remnants remain highly
uncertain (Dunne et al.\ 2003; 2009; Krause et al.\ 2004; Barlow et
al.\ 2010; Matsuura et al.\ 2011).  Observations of historical
Galactic remnants are important since these are (1) resolved, so that
the different SN and interstellar/circumstellar tracers can be
separated, and (2) young enough to ensure the thermal emission is not
dominated by swept-up material. The Crab Nebula is one of only a few
sources which satisfy these criteria and was chosen to be observed as
part of the \Hersc~guaranteed time project Mass Loss from Evolved
StarS (Groenewegen et al.\ 2011).  This survey includes Cas A
(Barlow et al.\ 2010) and the Type Ia remnants {\it Kepler} and {\it Tycho} (Gomez
et al.\ 2012; see also Morgan et al.\ 2003 and Gomez et al.\ 2009).
Unlike these remnants, the Crab has
negligible cirrus contamination along the line of sight and is an
ideal source to minimize the effects of unrelated interstellar
material.

The Crab Nebula has been an object of interest for a number of years
(see Hester 2008 for a comprehensive review). The remnant of an
explosion in 1054\,AD, the Crab is a pulsar wind nebula lying at a
distance of 2\,kpc (Trimble 1968). Its structure can be separated into
 two major components: the pulsar wind nebula (seen in X-rays and
optical) with smooth synchrotron emission (at near-IR and radio
wavelengths), and a network of filaments (traced in the optical and
IR) consisting of thermal ejecta. The low expansion velocity of the
ejecta suggests the remnant is the result of a Type II-P explosion
(MacAlpine \& Satterfield 2008). This is further confirmed by
abundance constraints which put the progenitor star at $9$-$12\,
M_{\odot}$ (Nomoto et al.\ 1982; Nomoto 1985; MacAlpine \& Satterfield
2008).

Unusually amongst supernova remnants (SNRs), the material in the Crab
is primarily photoionized by non-thermal radiation from the
synchrotron nebula (e.g. Davidson \& Fesen 1985; MacAlpine \&
Satterfield 2008).  The latter authors found the main nebular gas
component to be highly nitrogen-depleted and carbon-rich, although two
other gas components with C/O $< 1$ were also found to be present. The
implication of this is that both carbon-rich {\em and} oxygen-rich
dust species could exist in the nebula. Using optical data, Woltjer \&
V\'{e}ron-Cetty (1987) detected the presence of absorption
attributable to dust at the position of a bright [O\,{\sc iii}]
filament. Fesen \& Blair (1990) obtained high angular resolution
optical continuum images, revealing the presence of numerous ``dark
spots'' across the synchrotron nebula coincident with bright emission
cores seen in narrow-band [O\,{\sc i}], [C\,{\sc i}] and [S~{\sc ii}]
images, consistent with dust existing in partially ionized or neutral
clumps.

Previous IR studies confirmed the presence of dust grains in the Crab
as early as the 1980s with masses ranging from $0.005$ to $0.03\,M_{\odot}$
(e.g. Marsden et al. 1984).  Green et al. (2004) used {\it
  ISO} and SCUBA to infer the presence of $0.02$-$0.07\,
M_{\odot}$ of dust depending on its composition. A careful analysis of
the line contribution to the mid-IR using {\it Spitzer} data out to
70\,$\mu$m suggested a dust mass of $0.001$-$0.01\, M_{\odot}$ (Temim
et al. 2006). Using spatially resolved MIR spectroscopy across the
remnant, Temim et al. (2012) later revised this to a silicate grain
mass of $(2.4^{+3.2}_{-1.2}) \times 10^{-3}\,M_{\odot}$.  

Previous studies either lacked long wavelength spectroscopic
information or adequate sampling of the FIR emission.  In this paper,
we present detailed FIR and crucially,
submillimeter-through-to-millimeter observations of the Crab Nebula
obtained with {\it WISE}, {\it Spitzer}, {\it ISO}, \Hersc~and {\it Planck}
which allow us to accurately determine the synchrotron contribution
and line emission to beyond 600\,$\mu \rm m$.  We can then estimate
the {\em total} dust mass within the ejecta for the first time.

\begin{table}
 \centering
 \begin{tabular}{@{} llcc @{}}
   \hline
   Inst.& RA, Decl. (J2000) & TDT/ObsID & Int. Time\\
\hline
\multicolumn{4}{c}{Photometric Imaging} \\
   \hline
SPIRE & $\rm 05^h34^m31^s.97$, $22^{\circ}00^{\prime}52^{\prime \prime}.10$ & 1342191181& 4555\,s \\
PACS & $\rm 05^h34^m31^s.97$, $22^\circ00^{\prime}52^{\prime \prime}.10$ & 1342204441 & 1671\,s  \\
 PACS & $\rm 05^h 34^m 31^s.97$, $22^\circ 00^{\prime}
52^{\prime \prime}.10$  & 1342204442 & 1671\,s \\
PACS &$\rm 05^h 34^m 31^s.97$, $22^\circ00^{\prime}52^{\prime \prime}.10$  & 1342204443 & 1671\,s \\
PACS & $\rm 05^h 34^m 31^s.97$, $22^\circ00^{\prime} 52^{\prime \prime}.10$  & 1342204444 & 1671\,s \\
  \hline
 \multicolumn{4}{c}{Spectroscopy }\\
  \hline
  ISO-1& $\rm 05^h 34^m 34^s.27$, $22^\circ01^{\prime}02^{\prime \prime}.4$ & 69501241 & 1124\,s \\
  ISO-2 & $\rm 05^h 34^m 32^s.02$, $22^\circ02^{\prime}04^{\prime \prime}.6$ & 69301542 & 1126\,s \\
  ISO-3  & $\rm 05^h 34^m 29^s.31$, $22^\circ00^{\prime}37^{\prime \prime}.0$ & 69301543 & 1124\,s \\
  ISO-4 & $\rm 05^h 34^m 34^s.19$, $21^\circ59^{\prime}54^{\prime \prime}.7$ & 69301611 & 1630\,s \\
  FTS & $\rm 05^h 34^m 29^s.47$, $22^\circ00^{\prime}30^{\prime \prime}.4$ &1342204022 & 3476\,s \\
 IFU & $\rm 05^h 34^m 29^s.44$, $22^\circ00^{\prime}32^{\prime \prime}.52$ & 1342217847 & 2267\,s  \\
  IFU & $\rm 05^h 34^m 29^s.42$, $22^\circ00^{\prime}47^{\prime \prime}.17$  & 1342217847 & 1139\,s \\
   \hline
 \end{tabular}
 \caption{\footnotesize Observation Log for \Hersc~PACS and SPIRE
   Photometry and {\it ISO} LWS, \Hersc~PACS IFU, and SPIRE FTS
   Data Sets of the Crab Nebula.}
 \label{tab:label}
\end{table}

\section{Observations and data reduction}
\label{sec:data}

\begin{figure}
\begin{center}
{\includegraphics[trim=18mm 12mm 18mm
  8mm,clip=true,width=8.5cm]{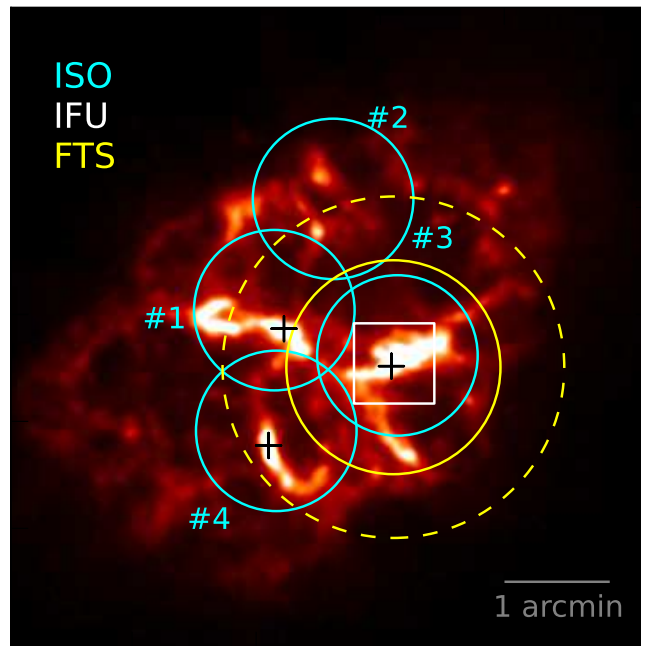}}
\end{center}
\figcaption{$3 \times 3$\,arcmin region showing the \Hersc~PACS image
  of the Crab Nebula at 70\,$\mu$m with
  the locations of the {\it ISO} LWS apertures (\#1-4 --small cyan circles), PACS
  IFU (white rectangle) and SPIRE FTS FoV (yellow dashed).  The unvignetted FTS
  FoV is also shown (solid yellow circle).  Black crosses mark the locations
  of the {\it Spitzer} spectra from Temim et
  al. (2012).  \label{fig:lines}}
\end{figure}

\subsection{{\it Herschel} Photometric Imaging}
\label{sec:data}

The Crab SNR was observed with the \Hersc~Photodetector
Array Camera (PACS; Poglitsch et al.\ 2010) and Spectral and
Photometric Imaging Receiver (SPIRE; Griffin et al. 2010) at 70, 100,
160, 250, 350 and 500\,$\mu$m (a summary of the observations is listed
in Table~\ref{tab:label}).  The PACS photometry data were obtained in
``scan-map'' mode with speed 20\,arcsec\,s$^{-1}$ including a pair
of orthogonal cross-scans over 22\,arcsec $\times$ 22\,arcsec.  In
order to obtain images at all three PACS wavelengths, we used both the
70+160\,$\mu$m and 100+160\,$\mu$m channels, leading to the 160-$\mu$m
image having twice the exposure time of the other two channels. The
SPIRE maps are ``Large Map'' mode with scan length of 30\,arcmin over
32\,arcmin $\times$ 32\,arcmin at a speed of 30\,arcsec\,s$^{-1}$; a
cross-scan is also taken, with a repetition factor of
three. The data were processed following the description given in
Groenewegen et al.\ (2011).

The PACS photometric data were reduced with the {\it Herschel}
Interactive Processing Environment (HIPE; Ott 2010) applying all
low-level reduction steps (including deglitching) to Level 1. The {\sc
  scanamorphos} software (Roussel 2012) was then used to remove
effects due to thermal drifts and uncorrelated $1/f$ noise of the
individual bolometers and create the Level 2 map.  The full width at half
maximum (FWHM) at 70, 100, and 160\,$\mu\rm m$ is 6, 8, and
12\,arcsec, respectively. The flux calibration uncertainty for PACS is
less than 10\% (Poglitsch et al.\ 2010) and the expected color
corrections are small compared to the calibration errors.  We
therefore adopt a 10\% calibration error.  

For SPIRE, the standard photometer pipeline (HIPE
v.5.0) was used (Griffin et al.\ 2010) with an additional iterative
baseline removal step (e.g. Bendo et al.\ 2010). The SPIRE maps were
created with the standard {\sc na\"ive} mapper (e.g. Griffin
et al. 2008).  We multiply the 350\,$\mu$m data product by 1.0067 to
be in line with the most recent calibration pipeline (v7). The FWHM
for pixel sizes of 6, 10, and 14\,arcsec is 18.1, 24.9, and 36.4\,arcsec at 250, 350 and 500\,$\mu\rm m$ respectively. The SPIRE
calibration methods and accuracies are outlined by Swinyard et al.\
(2010) and are estimated to be 7\%. The pipeline produces
monochromatic flux densities for point sources but at the longer
wavelengths, color corrections become significant and we therefore use
the correction factors listed in the SPIRE Observer's Manual (2011).

\subsection{{\it Herschel} PACS and SPIRE Spectroscopy}
\label{sec:specdata}

A 51-210-$\mu$m full-range spectral scan was obtained with the PACS
Integral Field Unit (IFU) Spectrometer (Poglitsch et al. 2010). The
IFU has 5$\times$5 spaxels with each spaxel being 9.4~arcsec on a
side. The coordinates (positioned on the brightest nebular filament -
Figure~\ref{fig:lines}), are listed in Table~\ref{tab:label}.  The
``chop-nod'' spectrometer mode was used, with the ``off'' position located
6~arcmin to the north and south of the ``on'' position for the red and
blue range scans, respectively. The ``chop-nodded'' observations
contained one single nodding cycle and one single up-down scan in
wavelength. The data were reduced to Level 2 products using the
standard PACS chopped large range scan and spectral energy
distribution (SED) pipeline in HIPE
version 8.0.1 (Ott 2010) using calibration file PACS\_CAL\_32\_0.

The SPIRE Fourier transform spectrometer (FTS) was used to obtain
sparse-map spectra. Two bolometer arrays provided overlapping bands
covering 32.0-51.5\,cm$^{-1}$ (194-313\,$\mu$m for the SPIRE Short Wavelength Spectrometer Array--SSW) and
14.9-33.0$\,\rm cm^{-1}$ (303-671$\,\rm \mu m$ for the SPIRE Long Wavelength Spectrometer Array--SLW). The SSW and SLW
beamsizes are $\sim$18 and $\sim$37\,arcsec, respectively. The two
yellow circles in Figure~\ref{fig:lines} show the 2.2-arcmin diameter
unvignetted and 3.2-arcmin diameter partially-vignetted field of view (FoV) for the
SPIRE spectrometer pointing. A point-source calibration was applied to the central detectors of each array
using Uranus (Swinyard et al.\ 2010).  As the Crab is
essentially fully extended in the beam, the calibration is
modified using the self-emission of the telescope as the primary
source (as we know both its temperature and emissivity properties).
This telescope Relative Spectral Response Function is
photometrically calibrated against Uranus using
knowledge of the instrument spatial response function as described in
the SPIRE Observers Manual (2011).  The resulting spectra have an
absolute calibration accuracy of 5\% compared to the Uranus flux model
of G. Orton et al. (in preparation). 

\subsection{{\it ISO} LWS Far-infrared Spectroscopy}
\label{isodata}

In addition to our {\em Herschel} spectroscopy, we have made use of
archival 43-197-$\mu \rm m$ spectra of the Crab obtained with the Long
Wavelength Spectrometer (LWS; Clegg et al. 1996) aboard {\it ISO}.
Figure~\ref{fig:lines} shows the positions of the four pointings of the
LWS superposed on the PACS 70-$\rm \mu m$ image of the Crab (cyan
circles -- Table~\ref{tab:label}).  The 44-110-$\rm \mu m$ parts of the
spectra were previously published in Green et al.\ (2004).  The data
used here have been processed through the {\it ISO}-LWS Highly
Processed Data Products pipeline (Gry et al. 2003) and further
modified by removing gain shifts between the individual detectors
using the central 100-$\rm \mu m$ detector as the reference.  The Crab
is relatively faint and the continuum spectrum beyond 150$\,\rm \mu m$
suffers from poor dark current removal and possible non-linear
response in the detectors due to thermal instabilities.  This portion
of the continuum is considered untrustworthy although the line fluxes,
in particular the [C\,{\sc ii}]~158-$\rm \mu m$ line, are well
calibrated.

\begin{figure*}
\begin{center}
{\includegraphics[trim=25mm 22mm 0mm
  2mm,clip=true,width=18cm]{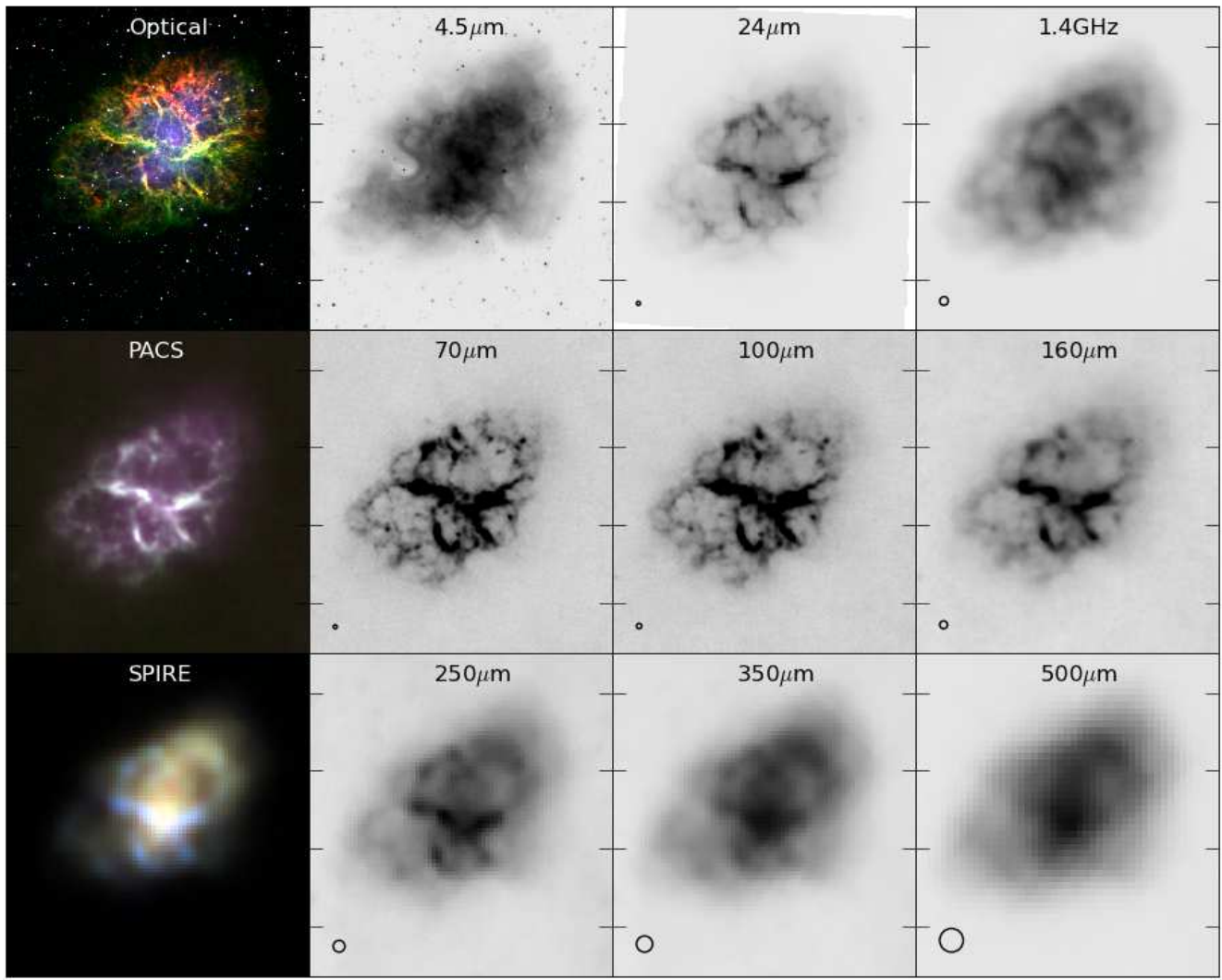}}
\end{center}
\figcaption{Multiwavelength montage of the Crab SNR centered on the
  pulsar ($\rm R.A. =262^{\circ}.671$, $\rm decl.=22^{\circ}.0145$,
  J2000.0) with diameter $4^{\prime}.2 \times 4^{\prime}.2$.  
    Top: (from left to right): three color image in H$\alpha$ (red),
  [O\,{\sc iii}] (green) and Bessel $B$ (blue) using images obtained
  from the Faulkes Telescope North (Appendix~\ref{sec:other}).  {\it
    Spitzer} IRAC 4.5\,$\mu$m, MIPS 24\,$\mu$m and VLA archive data at
  1.4\,GHz (taken in 1996). Middle: \Hersc~PACS three color
  image and individually the 70, 100 and 160\,$\mu$m images. 
    Bottom: \Hersc~SPIRE three color image and individually, 250, 350
  and 500\,$\mu$m images.  Beam sizes
  are indicated with the black circle in the lower left corner.  \label{fig:multi}}
\end{figure*}

\hspace*{0.4cm}

The global integrated fluxes at the \Hersc~wavelengths are
  measured using an elliptical aperture
  with size $245 \times 163^{\prime \prime}$ centered on the pulsar
  with the background contribution estimated using apertures
  off the remnant.
  These fluxes are
  combined with IR and submm fluxes from the literature as described
  in detail in Appendix~\ref{sec:other} with photometric fluxes from
  3.4 to 10,000\,$\mu$m listed in Table~\ref{tab:fluxes}
  (see also Figures~\ref{fig:specsed} and \ref{fig:sed}).  These are
  combined with our own flux measurements at 3.4-22\,$\mu$m with {\it WISE}
  and
  3.6-70\,$\mu$m
  {\it Spitzer}
  data (Appendix~\ref{sec:other}). 

When comparing the IR fluxes measured in this work (marked as ``A''
  in Table~\ref{tab:fluxes}) with the literature, we find that our PACS
  70\,$\mu$m flux for the Crab Nebula is significantly larger than the
  {\it Spitzer} measurement at the same wavelength (Temim et al.\
  2006).  At first glance this appears to suggest a huge (40\%)
  discrepancy between PACS and MIPS measurements (e.g., Aniano et al.\
  2012).  However, we find this discrepancy disappears if we use the
  most recent calibration factors from Gordon et al.\ (2007).
  Re-reducing the 70\,$\mu$m map using the most recent DAT instrument
  team pipeline produces a 70\,$\mu$m flux of
  210\,Jy after the relevant
  color corrections are applied, this is in excellent agreement
  with our PACS measurement.  Note that claimed differences between
  PACS and {\it Spitzer} MIPS can be resolved if instrumental effects
  are carefully considered; the PACS detectors are extremely
  stable with virtually no non-linearities compared to the Ge:Ga
  photoconductors employed in MIPS and indeed {\it ISO}. For these
  reasons, PACS has
  achieved an absolute calibration
  accuracy of 5\% for point sources with $<10\%$ quoted here (compared
  to $~20\%$ for MIPS at 70\,$\mu$m).

Since the radio
  synchrotron flux decreases with time by $-0.167 \,\% \rm \,yr^{-1}$
  (Aller \& Reynolds 1985),
  radio fluxes measured
  at different epochs (Table~\ref{tab:fluxes}) need to be
  corrected to the 2010 epoch to allow for comparison with the {\it WISE},
  \Hersc, and {\it Planck} data.   Note that this ``fading rate'' assumes that the
  non-thermal fluxes decline at the same rate at all wavelengths.

\section{Disentangling the contributions from different components}

In Figure~\ref{fig:multi}, we present a multiwavelength view centered on
the Crab, comparing the {\it Herschel} PACS and SPIRE images with
synchrotron emission seen at near-IR (with {\it Spitzer} IRAC) and at
radio wavelengths with the Very Large Array (VLA), and the ionized gas
observed at optical wavelengths.  The smooth non-thermal synchrotron
emission seen at 4.5\,$\mu$m appears to be confined within the
filamentary structures seen in optical and IR wavebands which
originate from ionized gas and dust emission.  Previous works have
shown that the warm dust component (seen in emission at 24\,$\mu$m)
traces the densest gas in the cores of filaments in low ionization
states (e.g., [O~{\sc i}] - Blair et al. 1997; Loll et al 2007) and this
is confirmed by the filamentary emission seen in the MIPS and
\Hersc~PACS images.  The distribution of the emission in the
\Hersc~SPIRE 250\,$\mu$m image is similar to the shorter IR
wavelengths.  At the longer SPIRE wavelengths (Figure~\ref{fig:multi}),
we start to see a strong resemblance with the radio emission at
1.4\,GHz, which traces both the smooth synchrotron seen at the shorter
4.5\,$\mu$m band, but also some filamentary emission arising from
free-free emission (e.g. Temim et al. 2006).  While the free-free
emission only makes a negligible contribution to the FIR continuum for
the Crab (see Appendix~\ref{sec:free}), the synchrotron component and
line emission are important at these wavelengths and need to be
removed before we can investigate if there is residual emission from dust.

\subsection{Line Emission}
\label{sec:lines}
\begin{figure*}
\begin{center}
{\includegraphics[trim=0mm 0mm 0mm
  0mm,clip=true,width=18cm]{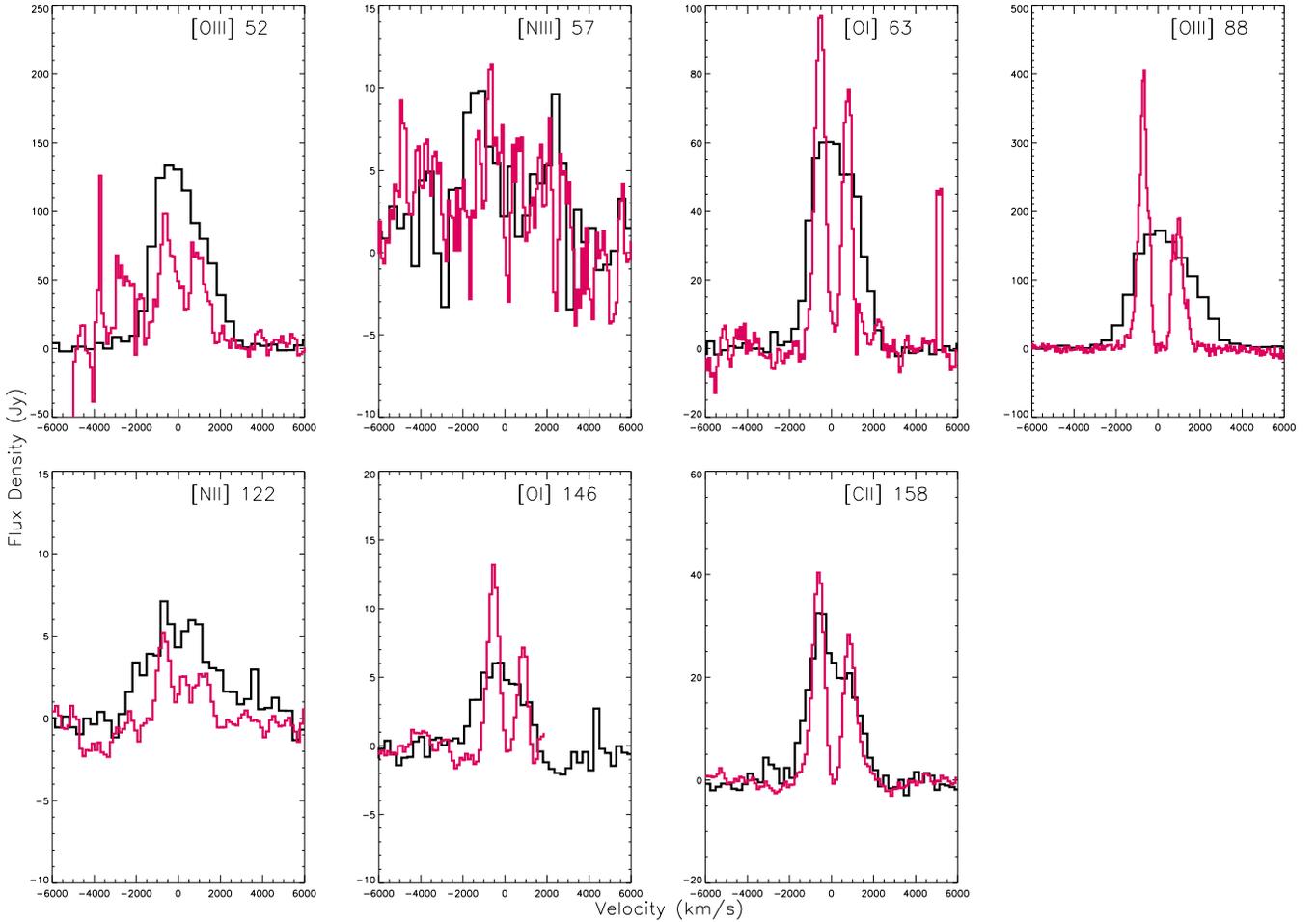}}
\end{center}
\figcaption{Velocity profiles of the emission lines detected in
  the Crab Nebula using the coadded PACS spectra (red) and the LWS
  spectrum (black) taken at the same location (\#3). The text in the top right
  corner lists the approximate central wavelength of the lines which
  at $51.8145\,\mu$m and 88.3564\,$\mu$m for [O~{\sc iii}] (NIST
  Atomic Spectra Database); 63.18371$\mu$m and 145.5255\,$\mu$m for
  [O~{\sc i}] (Zink et al.\ 1991); 157.74095\,$\mu$m for [C~{\sc ii}]
  (Cooksy et al.\ 1986); and 121.8976\,$\mu$m and 205.1783\,$\mu$m for
  [N~{\sc ii}] (Brown et al.\ 1994).  \label{fig:profile}}
\end{figure*}

Temim et al. (2012) present a comprehensive analysis of the MIR
spectral lines across the Crab Nebula, with a number of forbidden
lines identified up to 36\,$\mu$m.  They estimate that the
contribution of line emission to the 24\,$\mu$m broadband flux is 27\%,
54\%, and 48\% measured at different locations across the remnant (shown
in Figure~\ref{fig:lines}), suggesting on average, that line emission
contributes 43\% $\pm$ 6\% of the flux (Table~\ref{tab:lineconttest}).  (The
error quoted in this average is simply the range of values obtained
on the dense filaments.)  Note that this estimate of the line contribution was
obtained {\em after} Temim et al.\ subtracted synchrotron emission from the
{\it Spitzer} map.

\begin{table}[h]
\centering
\begin{tabular}{lccccc} \hline
\multicolumn{1}{c}{$\lambda$} & \multicolumn{5}{c}{Line Contribution
  in Band (\%)}  \\
 \multicolumn{1}{c}{($\mu$m)} & \multicolumn{1}{c}{}&
 \multicolumn{1}{c}{}& \multicolumn{1}{c}{}& \multicolumn{1}{c}{} &
 \multicolumn{1}{c}{Average}\\ \hline
24 & ... & 54&48  & 27  &  43 $\pm$ 6$^{\rm a}$ \\ \hline
&  \#1 &  \#2 &  \#3 &  \#4 &\\
70 & 6.6 & 4.2 & 5.4 & 4.4 & 4.90  $\pm$ 0.05$^{\rm b}$ \\
100 & 12.9& 5.6 & 9.6 & 6.7 & 8.7 $\pm$ 0.3$^{\rm b}$\\ \hline 
\end{tabular}
\tabcaption{\footnotesize The contribution to the total integrated
  flux at 24, 70 and 100\,$\mu$m at different locations across the
  remnant (Figure~\ref{fig:lines}).  $\rm ^a$ - This
  represents the contribution of line emission to the flux after synchrotron
  subtraction has been carried out (Temim et al.\ 2012).  $\rm ^b$ - The
  contribution at 70 and 100\,$\mu$m from line emission estimated using {\it ISO}
  LWS and \Hersc~PACS. \label{tab:lineconttest}}
\end{table}

In order to determine the contribution of line emission to the broad
band infrared fluxes of the Crab Nebula, \Hersc~PACS and SPIRE, and
{\it ISO} LWS data were analyzed.  Figure~\ref{fig:profile} shows the
velocity profiles of the seven emission lines detected in the co-added
\Hersc~PACS spectra and in the LWS spectrum taken at the same
location. Although the LWS line profiles are significantly broadened
compared to the instrumental resolution, it requires the higher PACS
spectral resolutions to fully resolve the individual blueshifted and
redshifted velocity components from the ejecta.  The separations
between the blue and red component emission peaks vary between 1290
and 1750$\rm \,km\,s^{-1}$, depending on the species (the profiles
will be discussed further in a separate paper).

The seven emission lines and their gas properties are discussed fully
in Appendix~\ref{sec:linedata} with fluxes measured relative to the
[O\,{\sc iii}] line provided in Table~\ref{tab:lineflux}.  These lines
suggest the Crab Nebula ejecta is nitrogen-depleted and carbon-rich
(unlike Cas A; Rho et al.\ 2008) implying that the dust is likely to
be carbon-rich.  However, there is also evidence that some regions in the
Nebula have solar-like CNO abundances which may allow silicate-type
grains to also form (see Appendix~\ref{sec:linedata} for more
details).

The four LWS spectra were used to estimate the contributions from the
emission lines to the measured PACS 70 and 100\,$\mu\rm m$ broad-band
fluxes.  To do this, the PACS filter spectral response functions were
convolved with and integrated across (1) the observed LWS spectra,
including emission lines; and (2) the LWS spectra after excising
emission lines and interpolating the adjacent continua across the
positions of the lines. The mean ratio of the case (2) to case (1)
in-band fluxes was found to be 0.951$\pm$0.012 for the 70\,$\mu \rm m$
filter and 0.913$\pm$0.028 for the 100\,$\mu$m filter
(Table~\ref{tab:lineconttest}).  We have applied these average line
correction factors to the measured 70- and 100-$\mu$m fluxes to obtain
the continuum-only broadband fluxes. The 146 and
158\,$\mu \rm m$ emission lines (Table~\ref{tab:lineflux}) contribute a
negligible amount to the broad-band PACS 160\,$\mu \rm m$ flux
measurement. No emission lines were detected in any of the spectra
from the SPIRE FTS detectors, the contribution of line emission to the
broad-band SPIRE photometry is therefore negligible.

\subsection{Synchrotron Emission}
\label{sec:synch}

In this section, we now estimate the contribution from the non-thermal
synchrotron component.  To do this, we fit a power law to the {\it
  Spitzer IRAC}, {\it WISE} and {\it Planck} data (large circles in
Figure~\ref{fig:sed}).

\begin{figure}
\begin{center}
{\includegraphics[trim=9mm 5mm 17.5mm
  0mm,clip=true,width=8.5cm]{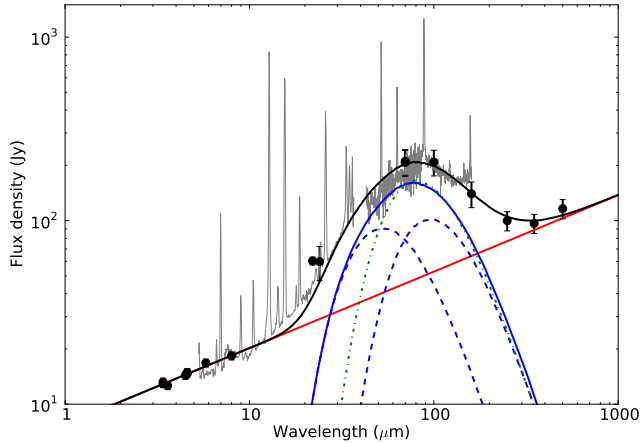}}
\end{center}
\figcaption{IR SED of the Crab Nebula including the integrated
  fluxes from \Hersc, the new {\it Spitzer} calibration and {\it WISE}
  fluxes (black points).  The errors include the photometric errors
  (dominated by calibration errors, Table~\ref{tab:fluxes}), the error
  in the synchrotron extrapolation and the error in the line emission
  contribution (Table~\ref{tab:lineconttest}) added in quadrature.
  The {\it ISO} \#3 spectra scaled to the 100\,$\mu$m flux is
  overplotted in gray.  The average (dereddened -- see Indebetouw et
  al.\ 2005) {\it Spitzer} spectroscopy (from the two brightest
  filaments in Figure~\ref{fig:lines}) is overplotted for comparison.
  The synchrotron power law is shown in red.  Two temperature
  components of amorphous carbon dust at 63 and 34\,K required to fit
  the SED are plotted (blue dashed) with the sum of these plotted in
  solid blue.  A single-temperature amorphous carbon fit (40\,K) to the FIR
  is also shown (green dot-dashed).  The sum of the two-component dust
  and synchrotron contributions is shown by the solid black curve.\label{fig:specsed}}
\end{figure}

\begin{figure}
\begin{center}
{\includegraphics[trim=7mm 11mm 26mm 0mm,clip=true,width=8.5cm]{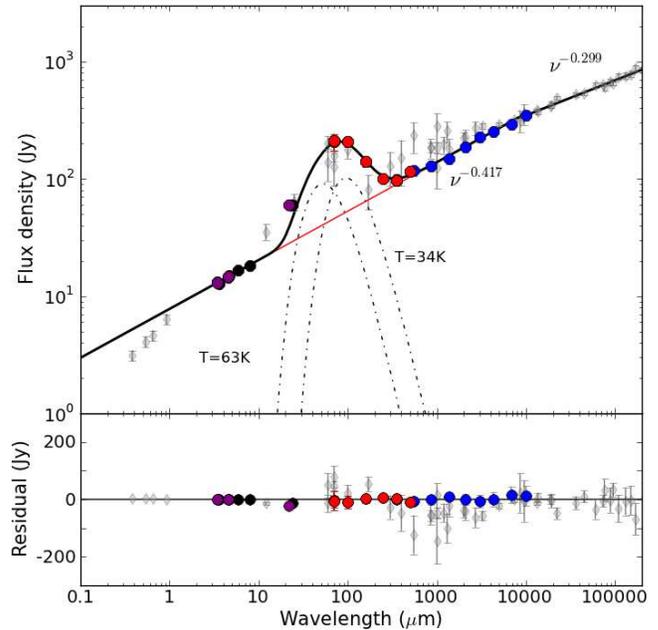}}
\end{center}
\figcaption{SED of the Crab Nebula from the IR-radio including
  \Hersc~(red points) and {\it Spitzer} (black), {\it WISE} (purple) photometry and {\it Planck} fluxes (Planck
  Collaboration 2011, blue points).  Previous fluxes from the
  literature (see Table~\ref{tab:fluxes} and references therein) are
  shown with gray diamonds.  The synchrotron law fitted to the
  $3.6-10^4\,\mu$m data points is the dashed black line.  Dot-dashed
  lines are the fitted components from thermal emission by amorphous
  carbon grains (see Figure~\ref{fig:specsed}). The solid black line is
  the total flux obtained from summing the synchrotron and the two
  dust components (with the residual shown below).  Note the total
  integrated fluxes plotted here also include a contribution from line
  emission at 24, 70, and 100\,$\mu$m (Section~\ref{sec:lines}), which has
  not been added to the total black SED curve plotted here.  \label{fig:sed}}
\end{figure}

Including the {\it Planck} fluxes in the SED allows us to fully
constrain the synchrotron power law between 3.4\,$\mu$m and 1\,cm for the
first time, using observations taken at the {\em same epoch}. A
least-squares fit to the {\it Spitzer}-{\it WISE}-\Hersc-{\it Planck}
data set produces a power law with frequency dependence $\nu^{-0.417}$
and amplitude 1489\,Jy at 1\,GHz (Figure~\ref{fig:sed}).  The error
  in the synchrotron spectral index ($\alpha$) from the line of best fit is
  $\pm 0.006$.  Extrapolating the expected fluxes at the
IR-submm wavebands due to synchrotron emission using the integrated
fluxes for the remnant with the above power law, we estimate the
synchrotron flux in each waveband (Table~\ref{tab:fluxes}).

The synchrotron power law in the FIR regime derived in this work is
steeper than the slope determined in the review by
Mac\'{i}as-P\'{e}rez et al. (2010), where the extrapolation to the
submm from the longest radio wavelengths suggested one synchrotron
component with $\alpha = -0.3$. We argue here that the exquisite
coverage between the IR and radio regime given by \Hersc~SPIRE and
{\it Planck} suggests that the synchrotron emission is described by
(at least) two power laws, with wavelengths beyond $10^4$\,$\mu$m
(30\,GHz), following the flatter relationship $\nu^{-0.3}$ (as also
seen in Green et al.\ 2004; Arendt et al.\ 2011). The slope then
breaks, becoming the steeper $\nu^{-0.42}$ law we find here.  In
  fitting this synchrotron power law, we do not incorporate the
  previous literature data obtained in the
  submillimeter-millimeter regime (shown by the faint gray diamonds in
  Figure~\ref{fig:sed}) since these
  fluxes often
  have large calibration errors, with small FoVs.  One
  should be particularly cautious when using the ISOPHOT data of the
  Crab (Green et al.\ 2004) since this was obtained in P32
  chopping mode, which can suffer significantly from transient effects
  with unreliable calibration (U. Klaas, private communication).  This
  dataset was never released as a scientifically validated measurement
  and therefore the quoted calibration accuracy for ISOPHOT (30\%) is
  not applicable for this dataset.   The previous literature
  measurements were also taken at different
  epochs and corrected using the average ``fading'' rate
  (Appendix~\ref{sec:other}).  With {\it WISE}, \Hersc, and {\it Planck}, not
  only are the photometric errors less than 10\%, the images also have a
  large background area to sample and the data were taken at the same
  epoch, therefore not relying on the application of a 
  correction factor.  

It is possible that the synchrotron spectrum could further break
  into different components in
  the IR regime (as suggested by Arendt et al.\ 2011) which
  would introduce further errors on the amount of synchrotron
  estimated at each wavelength.   However, we
  see no evidence for a sharp break either via
  the presence of excess
  flux or in differences in the spectral index maps created from the
  IRAC bands versus maps created from the
  VLA-\Hersc~images.   Unfortunately, given the low angular resolution
  in the FIR/submm, any local variations in the spectral index would
  not be seen in the method we have used in this work. The present
  data are therefore insufficient to separate out any small-scale
  variations in the synchrotron slope which may account for some of
  the residual emission. The shape of the SED at wavelengths beyond
  70\,$\mu$m and
  the {\it Planck} coverage
  rules out a significant break in the FIR/submm regime at least.
  However, we note that the flux
  attributed to synchrotron
  could be underestimated at 24\,$\mu$m if there is a break to a
  steeper power law at wavelengths less than 70\,$\mu$m.  

In order to spatially determine the distribution of synchrotron and
excess thermal emission from dust, we follow Temim et al. (2006) who
used a combination of the {\it Spitzer} IRAC and radio images to
obtain a spectral index map. We repeat this procedure using the
extinction-corrected 4.5\,$\mu$m IRAC image and the 500\,$\mu$m
\Hersc~SPIRE map, both of which are completely dominated by
synchrotron emission.  After aligning and convolving the IR-submm
images, we re-grid them to the same pixel size and deconvolve our
spectral index image with the appropriate beam.  We use the
deconvolved spectral index map to subtract the extrapolated
synchrotron flux expected at each IR-submm waveband using the spectral
index for that pixel.  In Figure~\ref{fig:spec} we show the extrapolated
synchrotron emission expected at 24\,$\mu$m using the spectral index
map made in this way.  Note that the subtraction of the synchrotron
component on a pixel-by-pixel basis removes the smooth emission at
24\,$\mu$m seen in Figure~\ref{fig:multi}, leaving only the excess
filamentary emission originating from the warm dust component.

\subsection{Thermal Emission from Dust}
\label{sec:mass}

Using the global SED of the Crab (Figures~\ref{fig:specsed} and
\ref{fig:sed}), after removing the contribution from line emission and
the well-constrained synchrotron component, we find the excess thermal
emission observed in the FIR can be described by the sum of
two modified blackbodies arising from a warm ($T_w$) and a cool
($T_c$) component.  Although we would expect a more complex temperature distribution for dust in the remnant,  a two-temperature component fit is adequate for a first-order approach in modeling the SED.  We can then fit the data with optical constants appropriate
for silicate or carbon grains, with the dust mass estimated using
Equation~(\ref{eq:dust}).  $S_{\nu}$ is the flux density, $D$ is the
distance, $B(\nu,T)$ is the Planck function and $\kappa_{\nu}$ is the
dust mass absorption coefficient calculated from the dust emissivity
$Q_{\nu}$ (for silicates -- Weingartner \& Draine 2001), and the grain
density $\rho$ (Laor \& Draine 1993):

\begin{equation}
M_d = {S_{\nu} D^2\over{\kappa_{\nu}} B(\nu,T)}.
\label{eq:dust}
\end{equation}

The total dust mass from the FIR model (see Table~\ref{tab:dust_mass})
for astronomical silicates is dominated by the cool component which
requires a best-fit temperature of $T_c=\rm 28.1^{+5.5}_{-2.8}\,K$ and
mass $M_d = 0.24^{+0.32}_{-0.08}\,M_{\odot}$. The warm component
(Figure~\ref{fig:specsed}) arises from $ (8.3^{+3.6}_{-6.4})\times
10^{-3}\,M_{\odot}$ of dust at a temperature of $\rm
55.6^{+7.8}_{-2.8} \,K$.

As the gas in the filaments is carbon-rich
(Appendix~\ref{sec:linedata}), the dust may be composed of amorphous
carbon grains. In this case (with $Q_{\nu}$ taken from Zubko et al.\
1996 (their `BE' model) and $\rho$ from Rouleau \& Martin 1991), the
dust mass arising from the cool component reduces to $0.11\pm
0.01\,M_{\odot}$ with $T_c = \rm 33.8^{+2.3}_{-1.8}\,K$. The warm
component requires only $M_d= (6.0^{+1.1}_{-2.4})\times
10^{-3}\,M_{\odot}$ at $T_w = 63.4^{+5.1}_{-2.7}\,\rm K$.  

We also attempted to fit iron grains to the SED (following Matsuura et
al.\ 2011) yet it is difficult to fit the MIR part of the thermal
emission.  To explain the FIR emission (i.e. beyond 70\,$\mu$m) with
iron grains, we require $ \sim 0.3\,M_{\odot}$ of dust at temperatures
34 and 69\,K with radius $0.1\,\mu$m (where $Q_{\nu}$ is taken from
Semenov et al.\ 2003 and $\rho$ from Nozawa et al.\ 2006).  However,
the mass of iron grains is highly dependent on the grain size, for
example grains with radius $<0.005\,\mu$m would require more than
$70\,M_{\odot}$ of dust.

We attempted to fit the residual emission from 24 to 350\,$\mu$m with
  a single-temperature modified blackbody, producing the parameters
  $M_d \sim 0.14\,M_{\odot}$ with $T\sim \rm 34\,K$ for astronomical
  silicates and $M_d \sim 0.08\, M_{\odot}$ with $T\sim \rm 40\,K$ for
  amorphous carbon (the green dot-dashed curve in
  Figure~\ref{fig:specsed}). However, the single component always
  severely underestimates the flux at 24\,$\mu$m. For this to be
  explained via contamination of line emission in this waveband, we
  would require 96\% of the broad-band flux at 24\,$\mu$m to be due to
  line contribution; this is clearly not supported by the careful
  analysis of IR spectra across the remnant in Temim et al.\ (2012).
  Given that the two-component model adequately fits the entire
  IR-submm SED and that the chi-squared statistic favors a
  two-component model fit (Table~\ref{tab:dust_mass}), we suggest this
  is the most valid model for
  the data. Ultimately, the difference in the final mass is small
  whether one or two dust populations exist if the grains are
  carbon-rich though the silicate dust mass estimated from the single-temperature fit is reduced by approximately half.  
  
\begin{table}
  \begin{tabular}{cccccc} \\\hline
& \multicolumn{5}{c}{Two-component Model}\\ 
    \multicolumn{1}{c}{} & \multicolumn{2}{c}{Warm Component} &
    \multicolumn{2}{c}{Cool Component} &\\
    \multicolumn{1}{c}{} & \multicolumn{1}{c}{$T_w$ (K)}  & $M_d$
    ($\times 10^{-3} M_{\odot}$) & \multicolumn{1}{c}{$T_c$ (K)} &
    $M_d$ ($M_{\odot}$) & $\chi^2$\\  \hline 
& & & && \\
Si & $55.6^{+7.8}_{-2.8}$ & $8.3^{+3.6}_{-6.4}$ &
$28.1^{+3.2}_{-5.5}$& $0.24^{+0.3}_{-0.1}$& 0.05\\
& & & & &\\
C & $63.4^{+5.1}_{-2.7}$ & $6.0^{+1.1}_{-2.4}$
&$33.8^{+2.3}_{-1.8}$  & $0.11\pm 0.01$ & 0.23\\ 
& & & & &\\ \hline
&\multicolumn{3}{c}{One-component Model} & & \\  
& \multicolumn{1}{c}{$T_d$ (K)}  & \multicolumn{1}{c}{$M_d$
  ($M_{\odot}$)} & $\chi^2$\\ \hline
& & & & &\\ 
Si & \multicolumn{1}{c}{34} & \multicolumn{1}{c}{0.14}  & $3\times 10^{-5}$ & & \\ 
C &  \multicolumn{1}{c}{40} & \multicolumn{1}{c}{0.08} &$7 \times
10^{-3}$ & & \\ \hline
\end{tabular}
\caption{\footnotesize Summary of the best-fit dust masses and temperatures for
  silicate and amorphous carbon grains using the two (top) and
    one (bottom) component dust
  models displayed in Figures~\ref{fig:specsed} and
  \ref{fig:sed}. Note: the reduced $\chi^2$ statistic is also included. \label{tab:dust_mass}}
\end{table}

Note that modeling the SED with dust grains with a continuous temperature distribution could produce a lower dust mass compared to that estimated using the two-temperature approach.   The dust masses quoted in Table~\ref{tab:dust_mass} should then be regarded as an upper limit on the mass of dust in the remnant.  However the largest uncertainty in a quoted dust mass arises from
the choice of optical constant.  This becomes more important when
choosing models within the ``umbrella'' term amorphous carbon, which
encompasses a number of different classes of materials such as soot,
glassy carbon (J\"{a}geret al.\ 1998), and the
ACAR/ACH2/BE models of amorphous grains from Zubko et al. (1996).
Although the optical constants vary depending on which of these
grain types is chosen, at FIR wavelengths $Q_{\nu}$ varies by only a
factor of a few for these different types (Figure~10 in
J\"{a}ger et al.\ 1998; Hanner 1988). One exception to this arises if
the grains are pyrolyzed cellulose where $Q_{\nu}$ is different by a
factor of 10 (J\"{a}ger et al. 1998). \\

\begin{figure}
\begin{center}
\includegraphics[trim=19.8mm 20mm 3mm
  0mm,clip=true,width=8.5cm]{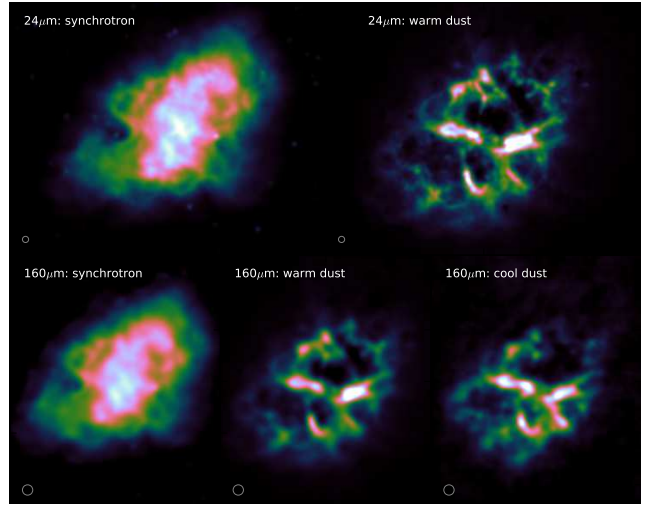}
\end{center}
\figcaption{Top: the distribution of synchrotron and thermal
  emission from the warm dust component at 24\,$\mu$m.  Bottom:
  the \Hersc~SPIRE 160\,$\mu$m emission separated into the synchrotron
  component (left), the warm dust component (as traced by the emission
  at 24\,$\mu$m) (middle) and the newly identified cool dust component
  (right).  The color palette used in
  the online version of this figure is the cubehelix scheme in which
  color monotonically increases in terms of perceived brightness
  (Green 2011).  \label{fig:spec}}
\end{figure}

The distribution of the warm dust component is shown in
Figure~\ref{fig:spec} using the synchrotron- and line
emission-subtracted 24\,$\mu$m image.  We can now use this warm dust
map to extrapolate the expected emission from the warm dust component
in the \Hersc~bands and reveal the distribution of the cool dust
component.  The spatial distribution of the synchrotron emission at
160\,$\mu$m (using the spectral index map) is shown in
Figure~\ref{fig:spec} along with the extrapolated warm and cool dust
components at this wavelength.  The morphology of this cool dust is
similar to the warm dust emission at 24\,$\mu$m and is distributed in
the dense filamentary structures as expected.

\section{Dust in the Crab Nebula}

Using {\it Spitzer} data out to 70\,$\mu$m, Temim et al.\ (2012)
derived a silicate grain mass of $2.4\times 10^{-3}\,M_{\odot}$ ($T
\sim 55\,\rm K$) and $3.2\times 10^{-3}\, M_{\odot}$ ($T \sim 60\,\rm
K$) for carbon grains. Their estimates are of similar order to the
mass of grains estimated from the hot component listed in
Table~\ref{tab:dust_mass}. However, the {\em total} dust mass derived
here is at least an order of magnitude higher than Temim et al.\ with
$M_d \sim 0.24\, M_{\odot}$ ($T_c \sim 28\,\rm K$) or
$0.11\,M_{\odot}$ ($T_c \sim 34\,\rm K$) for silicates and carbon
grains. The difference in mass between these two studies is not due to
the optical constants assumed, since Temim et al.\ use the ACAR model
from Zubko et al.\ (1996) which has similar $Q_{\nu}$ to the BE model
used here at FIR wavelengths.  The difference is due to the
  longer wavelength coverage of \Hersc~PACS and SPIRE, the matched-epoch
  observations and careful subtraction of synchrotron emission in the
  FIR--submillimeter regime.

To account for their dust temperature of $50$--$60\,\rm K$, Temim et al.\
(2012) modelled the dust heating (via non-thermal radiation) and
cooling rates (through IR emission) in the Nebula. They showed that
the IR emission ($\lambda<70\,\mu\rm m$) originates from small grains
with radii $< 0.015\,\mu$m. Their results imply that the SN grains in
the Crab are small, and therefore easier to destroy via sputtering;
the authors also point out that this is somewhat at odds with
theoretical models of SN grain formation (e.g. Kozasa et al.\
2009). Given the newly detected cool dust component in this work, the
long-wavelength radiation originates from larger grains with radius
$>0.06\, \mu$m.  Such grains are predicted by the Kozasa et al.\ model
and would suggest this new cool component of dust would be harder to
destroy.

Since the dust is spatially coincident with ionized ejecta material
(Figure~\ref{fig:profile}), it is plausible that the thermal emission
arises from newly formed grains.  Indeed, we can rule out a swept-up
interstellar origin using a simple argument (see, for example, the
{\it Tycho}
and {\it Kepler} remnants--Gomez et al.\ 2012).  The volume swept up by the
Crab is $\sim 15\,\rm pc^3$, sweeping up a total gas mass of
approximately one-tenth of a solar mass (Trimble 1970; Davidson \&
Fesen 1985) for typical interstellar densities.  Applying a standard
gas-to-dust ratio (Devereux \& Young 1990), the swept-up dust mass
would therefore be $<10^{-2}\,M_{\odot}$.  At the location of the Crab
(180\,pc above the Galactic plane), the interstellar density is
thought to be much less than the canonical value, indeed
the surrounding medium appears devoid of material (Davidson \& Fesen
1985), placing even more stringent constraints on the possibility that
the dust here is swept up material.

Are the dust masses estimated here sensible given the expected heavy
element abundances from the SN?  The total amount of heavy
elements (and therefore the mass available to form dust) in the SN
ejecta from a $9$--$12\,M_{\odot}$ progenitor star varies from
$0.2$ to $0.5\,M_{\odot}$ depending on the model used and the mass of the
progenitor (Maeder 1992; Woosley \& Weaver 1995; Limongi \& Chieffi
2003; Nomoto et al.\ 2006).  The limit on the abundances expected in
the ejecta from these theoretical models provides the constraint for
the maximum possible dust mass, these are 0.09\,$M_{\odot}$ for
carbon, 0.03\,$ M_{\odot}$ for MgO, and 0.04\,$M_{\odot}$ for $\rm
SiO_2$.  For silicate-rich grains, we can increase the available mass
to $ 0.19\,M_{\odot}$ if we allow iron to form grains, for example
including grain compositions such as $\rm FeSiO_3$.

From the elemental abundances constraint, both the silicate and carbon
dust masses estimated from the SED fitting are well within the
maximum possible dust masses allowed for these compositions.  Whether
the grains are silicate- or carbon-rich, the observed dust masses
suggest efficient condensation of metals in the filaments. 

\section{Conclusions and Discussion}
 \label{sec:conc}

 The combination of {\it Spitzer}, {\it WISE}, {\it ISO}, \Hersc, {\it Planck} photometry and
 spectroscopy at longer wavelengths than probed before, reveals a
 previously unknown cool dust component in the Crab Nebula located
 along the ionized filamentary structures.

 To reveal this new dust component, we carefully removed the
 synchrotron component using {\it WISE}, \Hersc~and {\it Planck} fluxes for
 the remnant measured in the same epoch.  We find a steeper
 $\nu^{-0.417}$ power-law variation with frequency for the non-thermal
 component which describes the emission from synchrotron at
 wavelengths of 3.4--10,000\,$\mu$m.

 The contribution from line emission is then removed using {\it Spitzer}
 IR spectroscopy (Temim et al.\ 2012) and our analysis of
 \Hersc~spectroscopy combined with {\it ISO} LWS archive data. FIR
 spectroscopy yields high O/N ratios and although appears to be
 carbon-rich, also has a component with solar-like CNO abundances. It seems
 likely that carbon and silicate grains could be located in the ejecta.

 The mass of dust estimated using the two-temperature approach to
 fitting the SED ranges from $0.1$ to $0.2\, M_{\odot}$ depending on
 whether the grains are composed of amorphous carbon or astronomical
 silicates respectively.  The warm and cool dust components are
 distributed in the filaments coinciding spatially with Doppler-shifted ejecta material (traced via spectroscopy).  This indicates
 the dust is spatially coincident with the ejecta material and not
 swept up circumstellar/interstellar material.

 Comparing with the expected elemental abundances in the ejecta, this
 work suggests the condensation of heavy elements into dust grains is
 efficient, and that the filaments may provide a viable environment to
 protect the dust from shocks.  The dust mass estimated for the Crab
 in this work is similar to the cool (unambiguously) associated SN
 dust mass observed in Cas A, and at the lower limit of the mass
 estimated in SN1987A using recent \Hersc~observations (Matsuura et
 al.\ 2011).  Todini \& Ferrara (2001) and Kozasa et al.\ (2009)
 predict between $0.1$ and $0.3\,M_{\odot}$ of dust should form in the
 ejecta from the (Type-IIP) explosions of progenitor stars with
 initial mass $<15\,M_{\odot}$, in agreement with the
 dust masses derived here.

 It is unclear how much of the newly formed ejecta dust will survive.
 Grains will be destroyed via thermal sputtering in the reverse shock
 due to collisions with electrons and/or ions, yet unlike the Cas A
 and SN1987A remnants, the Crab does not have a visible reverse shock
 today (Hester 2008).  The current environment in which the dust
 particles find themselves in appears relatively benign, as shown by
 the large amounts of molecular hydrogen comfortably surviving within
 the filaments (Graham et al.\ 1990; Loh et al.\
 2011). The dust particles in the Crab Nebula appear well set to
 survive their journey into the interstellar medium and contribute to
 the interstellar dust budget.  Future ALMA observations of this
 source and other remnants are crucial to disentangle synchrotron and
 dust emission on smaller scales.  This will be particularly important
 in comparing the time evolution of dust forming and being destroyed
 in remnants at different stages.

\begin{appendix}
\section{Ancillary Data}
\label{sec:other}

We used 3.6--70\,$\mu$m data from {\it Spitzer} IRAC and MIPS
(PI. R. Gerhz; Temim et al.\ 2006). The fluxes were corrected for
extended emission and color correction.  Calibration uncertainties
were assumed to be 5\% for IRAC and 10\% at 24 and 70\,$\mu$m and we
applied the most up-to-date calibration factors from Gordon et
al.\ (2007).

We obtained single exposure (level 1b) images of the {\it WISE} (Wright et
al. 2010) all-sky release through the NASA/IPAC Infrared Science
Archive at 3.4, 4.6 and 22\,$\mu$m. Montage was used to process and co-add
the single exposure images. Median filtering of the single exposure
frames was used to make basic cosmetic corrections. Calibration
factors were applied according to the {\it WISE} Explanatory Supplement with
photometric errors assumed to be 5\% (see e.g Mainzer et al.\ 2012).

For comparison purposes, we use optical images of the Crab
(Figure~\ref{fig:multi}) using the 2.0\,m Faulkes Telescope North on in
H$\alpha$ (200\,s) and [O\,{\sc iii}] narrowband filters (240\,s) and
the broad-band Bessel blue filter (200\,s).

Photometric fluxes measured in previous works were added to the
\Hersc~and {\it Spitzer} datasets presented here, including fluxes
from {\it {\it Planck}} (Planck Collaboration 2011), {\sc
  archeops} (Mac\'{i}as-P\'{e}rez et al.  2010), theKuiper
  Airborne Observatory (Wright et al. 1979), {\it IRAS} (Strom \&
Greidanus 1992), ISOPHOT and SCUBA (Green et al. 2004). The
fluxes listed in Table~\ref{tab:fluxes} that were not measured in this
work, were taken from the compilation of literature fluxes in
Mac\'{i}as-P\'{e}rez et al. (2010), Arendt et al.\ (2011), or from the
{\it Planck} archive.  In this work, we adopt calibration errors of
$15\%$ for {\it IRAS}.  For the ISOPHOT data, we can only assume a calibration
error of 30\% which is appropriate for scientifically valid datasets
(from mode P22, for example).  However, the Crab data have not been
scientifically validated so this is a lower limit on the flux error.

\begin{table}
\centering
 \begin{tabular}{@{} lllllll @{}} \hline 
$\lambda$& Epoch &$S_{\rm tot}$ &Error  & $S_{\rm synch}$ & Inst. & Ref. \\ 
($\mu$m) && (Jy) & (Jy) & (Jy) & &\\ \hline
3.4 & 2010 & 12.9 & 0.6 & 13.1 &{\it WISE} & {\it A}\\ 
3.6 & 2004 & 12.6& 0.22 & 13.2 &{\it Spitzer} & {\it a}\\ 
4.5 & 2004&14.4&0.26  & 14.5&{\it Spitzer}&{\it a} \\ 
4.6 & 2010 &14.7&0.75  & 14.6 &{\it WISE} &{\it A} \\ 
5.8 & 2004&16.8& 0.1 &16.1& {\it Spitzer}&{\it a} \\ 
8.0 & 2004&18.3 &0.13 &18.5&{\it Spitzer}&{\it a} \\ 
 22 & 2010 & 60.3 & 3.5 & 28.1 & {\it WISE} &{\it A} \\
24 & 2004& 59.8 & 6.0 &29.2&{\it Spitzer}&{\it a} \\
... & 2004& 59.3 & 5.9 &  ... & {\it Spitzer}&{\it A}\\
60 & 1998 & 140.7& 42.4&42.8& {\it ISO} &{\it b}\\     
...& 1983 & 210.0& 8.0&  ... &{\it  IRAS} & {\it c}\\      
70 & 2004&  208.0 & 33.3  & 45.6 & {\it Spitzer}&{\it A} \\ 
...& 2010& 212.8 & 21.3 & ... &\Hersc &{\it A}\\ 
100 & 1998 &128.2& 38.5 & 52.9 & {\it ISO} & {\it b} \\      
... & 1983 & 184.0& 13.0 & ... &{\it IRAS} &{\it c}\\      
 ...&2010& 215.2& 21.5 & ... & \Hersc & {\it A} \\
160 &2010& 141.8 & 14.2& 64.3& \Hersc &{\it A}\\ 
170 & 1998 &83.2& 26.5 & 66.0 &{\it ISO} & {\it b} \\    
250 &2010 &103.4& 7.2& 77.5 & \Hersc &{\it A}\\ 
300 & 1979 &135.0 &41.0& 83.6 & KAO & {\it d}\\
350 &2010 & 102.4&  7.2 & 89.2 & \Hersc & {\it A}\\   
... & 2010 & 99.3 & 2.4*  & ... & {\it Planck} &{\it f}\\ 
400 & 1979 &158.0 & 63.0 & 94.3& KAO &{\it d}\\
432 & 2007 & 224 & 24 & 97.4 & IRAM &{\it e}\\
500 &2010 & 129.0&  9.0 & 103.5 &\Hersc &{\it A}\\   
550 & 2010 & 117.7& 2.1*& 107.7 &{\it Planck} &{\it f}\\    
-& 2002 & 237.0 & 68.0 & ...& Archeops & {\it g}  \\ 
850 & 2002 & 186.0 & 34.0  & 129.0 & Archeops& {\it g}   \\ 
... & 2010 & 128.6 & 3.1* &  .. &{\it Planck}& {\it f}\\
...& 1999 &190.0 & 19.0& ...  & SCUBA & {\it b}\\      
1000 & 1979 &131.0 &42.0 & 138.1 & CH &{\it d}\\
... & 1983 &194.0 &19.0 & ... &Mt. Lemmon &{\it h}\\
... & 1976 &300.0 &80.0 &... &Hale&{\it i}\\
1382 & 2010& 147.2 & 3.1* & 158.1&{\it Planck}&  {\it f}\\
2098& 2010& 187.1& 2.0* & 188.1& {\it Planck}& {\it f}\\
3000 & 2010& 225.4& 1.1*&218.4& {\it Planck}& {\it f}\\
4286 & 2010& 253.6 & 2.5*&253.4& {\it Planck}& {\it f}\\
6818 &2010& 291.6& 1.3* &307.5& {\it Planck}&{\it f}\\
10000& 2010 & 348.2& 1.2* &360.7& {\it Planck}&  {\it f} \\ \hline
\end{tabular}
\caption{\footnotesize Integrated fluxes for the Crab Nebula.\\
{\bf Notes:}   Also
  included are the synchrotron fluxes derived from $\nu^{-0.417}$. \\  
{\bf References:} {\it A} - this work; {\it a} -
  Temim et al. (2006); {\it b} - Green et al. (2004); {\it c} - Strom
  \& Greidanus (1992); {\it d} - Wright et al. (1979); {\it e} - Arendt
  et al.\ (2011) {\it f} -
  {\it Planck} Collaboration 2011; {\it g} - Mac\'{i}as-P\'{e}rez et al
  (2010); {\it h} - Chini et al. (1984); {\it i} - Werner et
  al. (1977). * - These errors are the uncertainty in flux quoted in
  the {\it Planck} catalog which does not include calibration errors ($<7\%$).  CH - University of Chicago 
  photometer. KAO - Kuiper Airborne Observatory.} 
\label{tab:fluxes}
\end{table}

\section{Contribution from Free--Free Emission}
\label{sec:free}
Using a simple argument we can demonstrate that free--free radiation
makes a negligible contribution to the 200\,-$\mu$m continuum flux of
the Crab Nebula.

An upper limit can be derived starting from observed $F({\rm H}\beta)$
values (taken here to be $1.78\times 10^{-11} \rm \,erg\, cm^{-2}
\,s^{-1}$ from MacAlpine \& Uomoto 1991).  They also measure a global
$F(5876{\rm\AA})/F({\rm H}\beta) \sim 0.84$, while Davidson (1987) measured
$F(4686{\rm \AA})/F({\rm H}\beta) \sim 0.74$ and Smith (2003) measured
$\sim 0.72$ giving a mean ratio of 0.73.  De-reddening these ratios
with $E(B-V) = 0.47$, adopting a mean temperature ($T_e \sim 9000$\,K)
and using standard recombination emissivities, gives $n({\rm
  He}^+)/n({\rm H}^+) = 0.44$ and $n({\rm He}^{2+})/n({\rm H}^+) =
0.65$.

Using the above numbers with Equation 6 from Milne \& Aller (1975),
substituting in $F_{\nu} (5\,{\rm GHz})/I({\rm H}\beta)$, where $F_{\nu}
(5\,\rm GHz)$ is the 6\,cm free--free flux and $I({\rm H}\beta) = 8.6\times
10^{-11 } \,\rm erg\,cm^{-2}\, s^{-1}$ is the de-reddened $H\beta$
flux, predicts a value at 5\,GHz of $F_{\nu} = 0.88\,\rm Jy$.
Allowing for the $\nu^{-0.09}$ Gaunt factor frequency dependence of
free--free emission, the extrapolated flux at 200\,$\mu$m due to the
free--free component is $F_{\nu}({\rm ff}) = 0.53\,\rm Jy$ and 0.44\,Jy
at 24\,$\mu$m.

\section{Line emission, electron densities, and ionic abundances}
\label{sec:linedata}

\begin{table*}
 \centering
 \begin{tabular}{@{} ccccccc @{}}
   \hline
    & Coadded LWS & LWS \#1 & LWS \#2 & LWS \#3 & LWS \#4 & Coadded PACS \\
   \hline

[O {\sc iii}] 52 $\mu$m  & 116 $\pm$ 9 & 168 $\pm$ 12 & 109 $\pm$ 2
& 140 $\pm$ 9 & 91 $\pm$ 4 & (75 $\pm$ 4)$^{\rm a}$ \\

[N {\sc iii}] 57 $\mu$m & 11 $\pm$ 2 & 14 $\pm$ 2 & 5 $\pm$ 1 & 18 $\pm$ 4
&  &  \\

[O {\sc i}] 63 $\mu$m & 46 $\pm$ 2 & 69 $\pm$ 3 & 39 $\pm$ 1 & 58 $\pm$ 2 &
36 $\pm$ 2 & 46 $\pm$ 2 \\

[O {\sc iii}] 88 $\mu$m & 100 & 100 & 100 & 100 & 100 & 100 \\

[N {\sc ii}] 122 $\mu$m & 3.4 $\pm$ 0.2 & 8 $\pm$ 1 & 2.5 $\pm$ 0.3 & 2.9
$\pm$ 0.3 & 2.6 $\pm$ 0.3 & 2.0 $\pm$ 0.2 \\

[O {\sc i}] 146 $\mu$m & 2.2 $\pm$ 0.1 & 4.0 $\pm$ 0.2 & 1.9 $\pm$ 0.1 &
1.5 $\pm$ 0.3 & 1.9 $\pm$ 0.3 & 2.8 $\pm$ 0.2 \\

[C {\sc ii}] 158 $\mu$m & 7.8 $\pm$ 0.3 & 12.4 $\pm$ 0.8 & 6.9 $\pm$ 0.4 &
12 $\pm$ 1 & 5.9 $\pm$ 0.3 & 9.3 $\pm$ 0.5 \\

$F_{(88~\mu \mathrm{m})}$ ($\times 10^{-14}$ W m$^{-2}$) & 14.9 $\pm$ 0.4
& 1.94 $\pm$ 0.08 & 5.9 $\pm$ 0.1& 2.19 $\pm$ 0.05 & 4.64 $\pm$ 0.09 & 3.6
$\pm$ 0.1 \\

&&&&&&\\

$F(52)/F(88)$ & 1.16 $\pm$ 0.07 & 1.68 $\pm$ 0.07 & 1.1 $\pm$ 0.03 & 1.4
$\pm$ 0.07 & 0.91 $\pm$ 0.05 & \\

$n_e$(O {\sc iii}) (cm$^{-3}$) & 240 $\pm$ 30 & 485 $\pm$ 30 & 220 $\pm$ 20
& 350 $\pm$ 30 & 135 $\pm$ 20 & \\
&&&&&&\\

$F(52+88)/F(57)$ & 18.5 $\pm$ 1 & 18.9 $\pm$ 0.7 & 42.8 $\pm$ 3 & 12.8 $\pm$
1 & & \\

O$^{2+}$/N$^{2+}$ & 15.0 $\pm$ 0.9 & 14.4 $\pm$ 0.5 & 35 $\pm$ 3 & 17 $\pm$
1 & & \\

&&&&&&\\

$F(122)/F(57)$ & 0.29 $\pm$ 0.05 & 0.54 $\pm$ 0.1 & 0.51 $\pm$ 0.1 & 0.16
$\pm$ 0.04 & & \\

N$^+$/N$^{2+}$ & 2.3 $\pm$  0.4& 5.3 $\pm$ 1& 3.9 $\pm$ 1 & 1.4 $\pm$ 0.4 &
& \\
   \hline
 \end{tabular}
 \caption{\footnotesize Far-infrared line fluxes for the Crab Nebula including electron densities and
   relative ion abundances of the gas.  The relative fluxes are given on
   a scale where $F(88\,\mu \rm m) =100$.  $^{rm a}$ The PACS relative flux calibration is uncertain at 52\,$\mu$m.}
 \label{tab:lineflux}
\end{table*}

Table~\ref{tab:lineflux} lists the integrated fluxes in the 88\,$\mu$m
[O\,{\sc iii}] line measured in each of the four LWS spectra, as well
as its flux in the co-added LWS spectrum and in the co-added spectra
from the 25 PACS IFU spaxels. The table also lists the relative fluxes
for the other detected lines (Figure~\ref{fig:profile}), on a scale
where $F(88\,\mu$m) = 100.0. The [N\,{\sc iii}] 57\,$\mu$m line is
weakly detected in the spectra. The [O\,{\sc iii}] 52\,$\mu$m line
shows a well-resolved double-peaked profile in the co-added PACS
spectrum; however as it lies at the extreme short-wavelength end of
the PACS wavelength coverage, where the responsivity is falling
steeply, its relative flux calibration is uncertain and so we rely on
the LWS measurements.

Since the observed [O\,{\sc i}] and [C\,{\sc ii}] lines can arise from
mainly, or partly, neutral regions, we confine our analysis here to
lines that originate from mainly ionized regions of the nebula.  The
flux ratio of the [O\,{\sc iii}] 52 and 88\,$\mu$m lines is an electron
density diagnostic (see Figure~3 of Liu et al. 2001).  We have used the
same O$^{2+}$ atomic datasets as Liu et al. to derive electron
densities of 135--485$\,\rm cm^{-3}$ from the LWS 52/88 line flux
ratios listed in Table~\ref{tab:lineflux}.  These values are somewhat
smaller than the values of 830--1230\,$\rm cm^{-3}$ derived from the
ratios of the [S\,{\sc ii}] 6716\,\AA\ and 6731\,\AA\ line fluxes
measured at several positions in the Nebula by MacAlpine et
al. (1996), and from the ratios of the [S\,{\sc iii}] 18.7 and
33--5$\,\rm \mu m$ line fluxes measured at several positions by Temim
et al.  (2012).  {\em HST} imagery presented by Sankrit et al. (1998)
has demonstrated that the optical [O\,{\sc iii}] emission from the Crab
originates from diffuse sheaths around the filamentary cores that are
bright in emission lines from lower ionization species, consistent
with the lower [O\,{\sc iii}] densities that we find here.

In Table~\ref{tab:lineflux} we present flux ratios and the resulting
derived ion ratios estimated using our tabulated $n_e$(O~{\sc iii})
values and the atomic data sets adopted by Liu et al.\ (2001).  Liu et
al. (2001) have shown that, because of their similar critical
densities, the ratio of the fluxes in the [O\,{\sc iii}]
(52+88)\,$\mu$m lines to that in the [N\,{\sc iii}] 57\,$\mu$m line
yields O$^{2+}$/N$^{2+}$ ion abundance ratios that are insensitive to
the adopted electron density. In addition, because of their similar
ionization potentials, O$^{2+}$/N$^{2+}$ is a good approximation to
O/N.  $F(52+88)/F(57)$ ratios measured from the LWS spectra, together
with the O$^{2+}$/N$^{2+}$ ion ratios derived from them are listed in
Table~\ref{tab:lineflux}.  The flux ratio $F(122)/F(57)$ and the
resulting derived N$^+$/N$^{2+}$ N$^+$/N$^{2+}$ ratios show singly
ionized nitrogen to be the dominant nitrogen ion in the Crab.  The
O$^{2+}$/N$^{2+}$ (O/N) ratios from the individual LWS spectra
straddle a range of 10--34 by number. These values can be compared with
the elemental abundances estimated for three `Domains' by MacAlpine \&
Satterfield (2008).  If we assume the solar abundances of Asplund et
al. (2009) then using their Table~2, their mass fractions correspond
to O/N number ratios of 7.2 (solar), 21 and 260 for Domains 1, 2, and
3, respectively. The individual O/N values derived here from LWS
spectra \#1--3 are consistent with Macalpine \& Satterfield's Domain
2. They found this to be the most prevalent material, with nitrogen
depleted by a factor of three and carbon enhanced by a factor of six,
relative to solar, which they noted to be consistent with a precursor
star $M_i \geq9.5~\, M_{\odot}$.

\end{appendix}

\acknowledgments We thank the referee Tea Temim for her constructive
and helpful referee report.  This research made use of APLpy, an
open-source plotting package for Python hosted at 
  http://aplpy.github.com. We thank Robbie Auld, Eli Dwek, Stuart Lowe, Takaya
Nozawa and Matt Smith for informative discussions and J D Armstrong
for support with the Faulkes Telescope North.  HLG acknowledges the
support of Las Cumbres Observatory.  This publication makes use of
data products from the {\it Wide-field Infrared Survey Explorer}, which is a
joint project of the University of California, Los Angeles, and the
Jet Propulsion Laboratory/California Institute of Technology, funded
by the National Aeronautics and Space Administration.  PACS has been
developed by a consortium of institutes led by MPE (Germany) and
including UVIE (Austria); KU Leuven, CSL, IMEC (Belgium); CEA, LAM
(France); MPIA (Germany); INAF- IFSI/OAA/OAP/OAT, LENS, SISSA (Italy);
IAC (Spain). This development has been supported by the funding
agencies BMVIT (Austria), ESA-PRODEX (Belgium), CEA/CNES (France), DLR
(Germany), ASI/INAF (Italy), and CICYT/MCYT (Spain). SPIRE has been
developed by a consortium of institutes led by Cardiff Univ. (UK) and
including: Univ. Lethbridge (Canada); NAOC (China); CEA, LAM (France);
IFSI, Univ. Padua (Italy); IAC (Spain); Stockholm Observatory
(Sweden); Imperial College London, RAL, UCL-MSSL, UKATC, Univ. Sussex
(UK); and Caltech, JPL, NHSC, Univ. Colorado (USA). This development
has been supported by national funding agencies: CSA (Canada); NAOC
(China); CEA, CNES, CNRS (France); ASI (Italy); MCINN (Spain); SNSB
(Sweden); STFC, UKSA (UK); and NASA (USA). HIPE is a joint development
by the \Hersc~Science Ground Segment Consortium, consisting of ESA,
the NASA \Hersc~Science Center and the HIFI, PACS and SPIRE consortia.

{\it Facilities:} \facility{Herschel (PACS and SPIRE)},
\facility{Spitzer}, \facility{Planck}, \facility{ISO}

\end{document}